\documentclass[twocolumn,secnumarabic, amsmath,amssymb]{revtex4-1}

\usepackage{graphicx}
\usepackage{bm}
\usepackage{color}

\begin{document}

\title{Mechanochemical subcellular-element model of crawling cells}%

\author{Mitsusuke Tarama$^1$}%
\email[]{mitsusuke.tarama@riken.jp}
\author{Kenji Mori$^2$}%
\author{Ryoichi Yamamoto$^{2,3}$}%

\affiliation{
$^1$%
Laboratory for Physical Biology, 
RIKEN Center for Biosystems Dynamics Research, 
Kobe 650-0047, Japan
}

\affiliation{
$^2$%
Department of Chemical Engineering, Kyoto University, Kyoto 615-8510, Japan
}

\affiliation{
$^3$%
Institute of Industrial Science, The University of Tokyo, Tokyo 153-8505, Japan
}

\date{\today}%

\begin{abstract}
Constructing physical models of living cells and tissues is an extremely challenging task because of the high complexities of both intra- and intercellular processes. In addition, the force that a single cell generates vanishes in total due to the law of action and reaction. The typical mechanics of cell crawling involve periodic changes in the cell shape and in the adhesion characteristics of the cell to the substrate. However, the basic physical mechanisms by which a single cell coordinates these processes cooperatively to achieve autonomous migration are not yet well understood. To obtain a clearer grasp of how the intracellular force is converted to directional motion, we develop a basic mechanochemical model of a crawling cell based on subcellular elements with the focus on the dependence of the protrusion and contraction as well as the adhesion and deadhesion processes on intracellular biochemical signals. By introducing reaction-diffusion equations that reproduce traveling waves of local chemical concentrations, we clarify that the chemical dependence of the cell-substrate adhesion dynamics determines the crawling direction and distance with one chemical wave. 
Finally, we also perform multipole analysis of the traction force to compare it with the experimental results. 
To our knowledge, our present work is the first study that accomplishes fully force-free migration utilizing intracellular chemical reactions. Although the detailed mechanisms of actual cells are far more complicated than our simple model, we believe that this mechanochemical model is a good prototype for more realistic models.
\end{abstract}

\maketitle

\section{Introduction} \label{introduction}

Cells are the basic units of living organisms. 
Despite the diversity of cellular types, they share similar structures. 
Biological cells are composed of a number of polymers, proteins, and lipids, which form stable structures such as the lipid membrane. 
At the same time, cells exist in a state far from equilibrium. 
Inside a cell, complex chemical reactions take place and are converted to mechanical forces by molecular motors. 
Consequently, biological cells spontaneously exhibit various dynamics, such as locomotion and proliferation. 

In general, objects that exhibit spontaneous motion are called active matter~\cite{Lauga2009The,Ramaswamy2010The,Vicsek2012Collective,Cates2012Diffusive,Marchetti2013Hydrodynamics,Bechinger2016Active}. 
In contrast to objects passively driven by external forcing, active matter, including living cells, generates force in itself, which is characterized by a vanishing force monopole due to the action-reaction law. 
Under this force-free condition, it is necessary to break symmetry to achieve spontaneous motion, such as directional motion. 
For microorganisms that swim in a fluidic environment, the scallop theorem~\cite{Purcell1977Life} describes the importance of breaking reciprocality to achieve net migration via internal cyclic motions. 

In nature, there exist a number of microorganisms that crawl on substrates, such as the extracellular matrix and other cells. 
In contrast to the locomotion of microswimmers, adhesion to the substrate plays an important role in the locomotion of crawling microorganisms. 
Such crawling motion is often observed in eukaryotic cells, such as Dictyostelium cells and keratocytes. 

Many models have been introduced to explain various aspects of the dynamics of crawling cells. 
Examples include the vertex model~\cite{Honda1978Description,Honda2004A,Honda2008Computer,Bi2016Motility-Driven}, 
the cellular Potts model~\cite{Nishimura2009Cortical,Niculescu2015Crawling}, 
the continuous model~\cite{Marchetti2013Hydrodynamics,Nier2016Inference}, 
the phase field model~\cite{Ziebert2016Computational,Shao2010Computational,Taniguchi2013Phase,Shi2013Interaction,Ziebert2012Model,Tjhung2015A}, 
and the particle-based model~\cite{Newman2007Modeling,Sandersius2008Modeling,Basan2011Dissipative,Zimmermann2016Contact,Smeets2016Emergent}. 
Here, we construct a model based on the particle-based model. 

In our previous study~\cite{Tarama2018Mechanics}, we systematically investigated the cycle of the typical crawling mechanism~\cite{Ananthakrishnan2007The}: 
1) protrusion at the leading edge,
2) adhesion to the substrate at the leading edge,
3) deadhesion from the substrate at the trailing edge, and
4) contraction at the trailing edge. 
To clarify the role of this cycle in efficient crawling motion, we introduced a simple mechanical model in which a cell is described by two subcellular elements connected by a viscoelastic bond, which includes an actuator that elongates and shrinks cyclically. 
The substrate friction of each element switches cyclically between the adhered stick state and the deadhered slip state. 
By tuning the phase shifts between the actuator elongation and the substrate friction of each element, we demonstrated that the order of the four basic processes of the typical crawling mechanism has a great impact on the crawling distance and efficiency, as well as the crawling direction. 

If we consider the extension of the mechanical model to a model cell consisting of far more than two elements or to two- or three-dimensional (2D or 3D) space, adjusting the phase shift of each element ``by hand'' becomes less feasible. 
Instead, we need to consider the underlying processes that regulate the actuator elongation and the stick-slip transition of the substrate friction. 

The aim of this paper is to develop a basic particle-based model for cell crawling. 
To this end, we consider a cell crawling on a flat substrate and extend our previous mechanical model to two dimensions. 
We describe a cell by a set of many subcellular elements connected by viscoelastic bonds~\cite{Newman2007Modeling}. 
In addition, intracellular chemical reactions are represented by simple reaction-diffusion (RD) equations~\cite{Murray2002Mathematical,Murray2003Mathematical,Kuramoto1984Chemical,Epstein1998An,Pismen2006Patterns}, which trigger mechanical activities. 
We then couple the RD equations and mechanical models to achieve efficient migration. 
In particular, we focus on the time delay between the intracellular chemical reactions and cell mechanics, which corresponds to the ordering of the basic crawling processes. 

This paper is organized as follows.
In the next section, we introduce the model equations that couples cell mechanics and intracellular chemistry. 
Then, we show that the dependence of the substrate adhesion on the intracellular chemistry determines the direction of the cell crawling in Sect.~\ref{sec:direct_retrograde_waves}.
In Sect.~\ref{sec:adhesion_mechanochemical}, we investigate how the cell crawling changes depending on the mechanical and chemical signals that control the substrate adhesion. 
The impact of the cell shape and size are studied in Sets.~\ref{sec:cell-shape} and \ref{sec:cell-size}, respectively. 
In Sect.~\ref{sec:random}, random crawling motion is realized by random excitation of intracellular chemistry, for which we analyze the traction force multipoles in Sect.~\ref{sec:traction}. 
Sect.~\ref{Summary and Discussion} is devoted to the summary and discussion, and this paper concludes with Sect.~\ref{sec:conclusion}.

\section{Model} \label{Model}

First, we introduce our mechanical model of a crawling cell and the RD equations representing intracellular chemical reactions. 
The choice of RD equations is arbitrary, and we employ a previously introduced model. 
We then couple the mechanical model and the RD equations, which regulate the intracellular mechanical activities. 
In particular, we confine ourselves to studying possible couplings between the intracellular chemical and mechanical models.

\begin{figure}[t]
\centering
\includegraphics{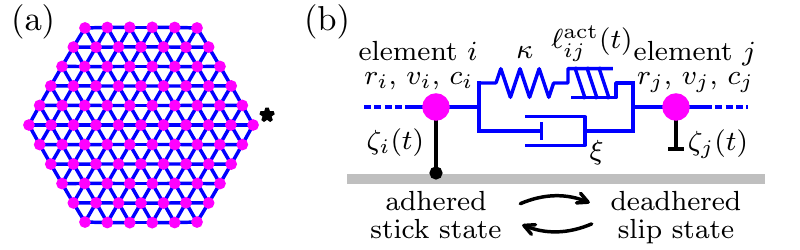}
\caption{
	Sketch of the subcellular element model of a cell crawling on a substrate. 
	(a) The cell is described by a set of subcellular elements (magenta circles) connected by viscoelastic bonds (blue lines). 
	The shape of a cell at rest is assumed to be a perfect hexagonal lattice. 
	The element indicated by the star is the activator element. 
	(b) Details of the subcellular elements and the connecting viscoelastic bond. 
	Each element possesses the chemical concentrations $\bm{c}_i$. 
	The actuator length $\ell^{\rm act}_{ij}(t)$ and the substrate friction coefficient $\zeta_i(t)$ change over time. 
}%
\label{fig:ScEM_schematics}
\end{figure}

\subsection{Subcellular-element model} \label{Subcellular-element model}

We describe a single cell by a set of subcellular elements~\cite{Newman2007Modeling} connected by Kelvin-Voigt type viscoelastic bonds, as schematically depicted in Fig.~\ref{fig:ScEM_schematics}. 
Since the typical size of a cell is on the order of ten micrometers, the effect of inertia is negligible. 
Then, the force balance equation of element $i$ is given by
\begin{align}
&\zeta_i(t) \bm{v}_i + \sum_{j \in \Omega_i} \xi \ell_{ij} ( \bm{v}_i -\bm{v}_j ) \notag \\
&\hspace{2em} = \sum_{j \in \Omega_i} \frac{\kappa}{\ell_{ij}} \hat{\bm{r}}_{ij} \left\{ r_{ij} - \left(\ell_{ij} +\ell^{\rm act}_{ij}(t) \right) \right\} +\bm{f}^{\rm area}_i,
\label{eq:force_balance}
\end{align}
where $\bm{v}_i$ is the velocity of the element $i$ located at the position $\bm{r}_i$. 
Here, the abbreviations $r = |\bm{r}|$ and $\hat{\bm{r}} = \bm{r}/r$ are used for the relative position $\bm{r}_{ij} = \bm{r}_j - \bm{r}_i$. 
The summation is over the set of the elements $\Omega_i$ connected to the element $i$ by the viscoelastic bonds. 

The first term on the left-hand side of Eq.~\eqref{eq:force_balance} represents the substrate friction with coefficient $\zeta_i(t)$, which changes over time due to intracellular activity. 
The second term represents intracellular dissipation with the rate $\xi$. 
The first term on the right-hand side represents intracellular elasticity with the elastic modulus $\kappa$ and free length $\ell_{ij}$. 
Intracellular activity is also included in the actuator, which tends to elongate the connecting bond by changing the free length over time as $\ell_{ij} +\ell^{\rm act}_{ij}(t)$. 
Here, $\ell^{\rm act}_{ij}(t)$ represents the actuator elongation, from which the force generated by the actuator is calculated as $\bm{f}^{\rm act}_{ij}(t) = -\kappa \ell^{\rm act}_{ij}(t) \hat{\bm{r}}_{ij} /\ell_{ij}$. 
We emphasize that the model Eq.~\eqref{eq:force_balance} satisfies the force-free condition since the intracellular force acts symmetrically on the pair of subcellular elements. 
Namely, the sum of the intracellular force in Eq.~\eqref{eq:force_balance} vanishes as
\begin{align}
&\sum_i \bm{f}^{\rm int}_i
 = 0.
 \label{eq:force-free}
\end{align}
where
\begin{align}
\bm{f}^{\rm int}_i
=&
- \sum_{j \in \Omega_i} \xi \ell_{ij} ( \bm{v}_i -\bm{v}_j ) \notag\\
&+\sum_{j \in \Omega_i} \frac{\kappa}{\ell_{ij}} \hat{\bm{r}}_{ij} \left\{ r_{ij} - \left(\ell_{ij} +\ell^{\rm act}_{ij}(t) \right) \right\} +\bm{f}^{\rm area}_i
 \label{eq:intracellular_force}
\end{align}
is the intracellular force acting on the element $i$.

The last term on the right-hand side of Eq.~\eqref{eq:force_balance} prevents the collapse of the subcellular element network, which is given by
$\bm{f}^{\rm area}_i = -\delta U^{\rm area} /\delta \bm{r}_i$, where $U^{\rm area} = \sum_{\langle i,j,k \rangle} \sigma/S_{ijk}^2$.
This potential $U^{\rm area}$ penalizes shrinking of the area of each triangle $\langle i,j,k \rangle$ formed by connected subcellular elements $i$, $j$, and $k$, which is defined by
$S_{ijk} = ( \bm{r}_{ij} \times \bm{r}_{ik} ) \cdot \hat{\bm{e}}_z /2$
with $\hat{\bm{e}}_z$ as the unit vector perpendicular to the 2D substrate. 

We scale the system by $L_0 = 10 \, \mu \rm{m}$ for length and $T_0 = 1 \, \rm{min}$ for time, which are physiologically relevant values for typical living cells ~\cite{Maeda2008Ordered,Tanimoto2014A}. 
In addition, the scale of the force is set to $F_0 = 10 \, {\rm nN}$, which is on the order of the traction force that cells exert on the substrate. 
The typical values of the mechanical parameters of the model Eq.~\eqref{eq:force_balance} are summarized in Appendix~\ref{sec:scaling}.

\subsection{Chemical reaction} \label{Chemical reaction}

In the model Eq.~\eqref{eq:force_balance}, the effects of the intracellular activities are included in the actuator elongation $\ell^{\rm act}_{ij}(t)$ and the change in the substrate friction coefficient $\zeta_i(t)$. 
The former represents the protrusion and contraction processes. 
The latter corresponds to the adhesion and deadhesion of the cell to the underlying substrate. 
In actual cells, such cellular activities are caused by various intracellular chemical signals.
However, it is not realistic to include all chemical components and their signaling pathways.
Therefore, we model the intracellular chemical reactions by simple RD equations. 

Here, we employ the RD equations proposed by Taniguchi et al.~\cite{Taniguchi2013Phase}, which are two-component activator-inhibitor equations: 
\begin{align}
\frac{\partial U_i}{\partial t} = D_U \nabla^2 U_i +G_U( U_i, V_i ),
\cr
\frac{\partial V_i}{\partial t} = D_V \nabla^2 V_i +G_V( U_i, V_i ),
\label{eq:Taniguchi_RD}
\end{align}
where $U_i$ and $V_i$ represent the inhibitor and activator concentrations for subcellular element $i$, respectively. 
In the original paper~\cite{Taniguchi2013Phase}, Eq.~\eqref{eq:Taniguchi_RD} were introduced to model the phosphoinositide signaling pathway of Dictyostelium cells, and the activator $U$ and the inhibitor $V$ corresponded to phosphatidylinositol (3,4,5)-trisphosphate (PIP3) and phosphatidylinositol (4,5)-bisphosphate (PIP2) concentrations, respectively.
The details of these RD equations are given in Appendix~\ref{sec:RD}. 

The important property of the RD equations, Eq.~\eqref{eq:Taniguchi_RD}, is that they are of the Grey-Scott type~\cite{Gray1983Autocatalytic,Gary1984Autocatalytic}. 
One of the advantages of the Grey-Scott model is that it can show either an excitable or a bistable nature depending on the parameters. 
Taniguchi et al. claimed that the signaling pathway that they were modeling was excitable, and thus, they considered the parameter region of the excitable case to successfully reproduce the experimental results~\cite{Taniguchi2013Phase}. 
Interestingly, similar RD equations were studied by Shao et al.~\cite{Shao2010Computational} in the context of cell crawling. 
However, they assumed a bistable regime to reproduce the steady migration of keratocyte cells. 

In this paper, we consider the excitable case with the parameters summarized in Table~\ref{table:parameters_RD}, following the study by Taniguchi et al.~\cite{Taniguchi2013Phase}. 
The Laplacian terms in Eq.~\eqref{eq:Taniguchi_RD} are calculated by using the moving particle semi-implicit (MPS) method~\cite{Koshizuka1998Numerical}. 
See also Appendix~\ref{sec:RD} for further information.

\subsection{Mechanochemical coupling} \label{Mechanochemical coupling}

To combine the cell mechanics, Eq.~\eqref{eq:force_balance}, and the RD equations, Eq.~\eqref{eq:Taniguchi_RD}, we consider the coupling of the chemical concentrations to the actuator elongation $\ell^{\rm act}_{ij}(t)$ and the substrate friction coefficient $\zeta_i(t)$ individually. 

Before introducing the coupling, we tested a specified traveling wave that couples to the actuator elongation and the substrate friction change. 
We introduced a time delay between the elongation and substrate friction change. 
The results showed that there exist an optimum time delay and optimum wavelength for which the cell exhibits the largest migration distance, as summarized in Appendix~\ref{sec:traveling_wave}.

\subsubsection{Actuator elongation} \label{sec:actuator_elongation}

First, we introduce the coupling between the RD equations and the actuator elongation. 
In Ref.~\cite{Taniguchi2013Phase}, actin polymerization was found to be enhanced with increasing PIP3 concentration. 
Therefore, we presume that the actuator elongation depends on the PIP3 concentration as
\begin{equation}
\ell^{\rm act}_{ij}( t ) = \ell_V \tanh [ a V_{ij}(t) ],
\label{eqTaniguchi_ell_ij}
\end{equation}
where $V_{ij}(t) = ( V_i(t)+V_j(t) )/2$ is the mean PIP3 concentration for the bond connecting the elements $i$ and $j$. 
Although $V_i(t)$ is a positive quantity, its maximal value depends on the strength of the initial fluctuation because of the excitable nature of Eq.~\eqref{eq:Taniguchi_RD}.
Therefore, $\tanh$ is introduced on the right hand side of Eq.~\eqref{eqTaniguchi_ell_ij} to prevent an extremely large elongation. 
$a$ is a constant denoting sensitivity, and $\ell_V$ is the magnitude of the elongation. 
Here, we set $a=\pi$ and $\ell_V=\ell_{ij}$.

\subsubsection{Substrate adhesion} \label{sec:adhesion}

\begin{figure}[t]
\centering
\includegraphics{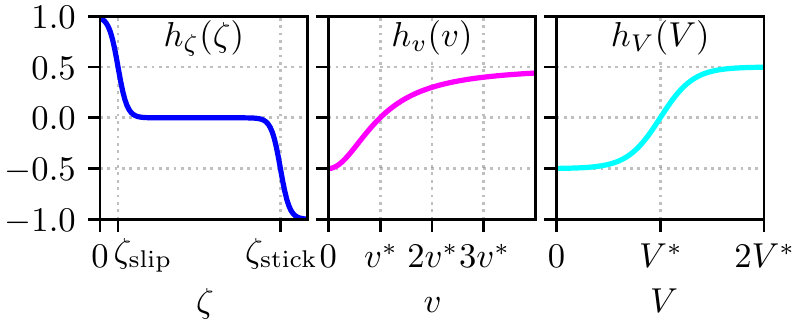}
\caption{
	Functional forms of $h_{\zeta}(\zeta)$, $h_v(v)$, and $h_V(V)$, which determine the substrate friction. 
	The parameters are set to the values given in Table~\ref{table:scales}.
}%
\label{fig:zeta_function}
\end{figure}

Next, we consider the adhesion to the substrate underneath and the deadhesion from it.
We model the adhesion/deadhesion processes by the transition of the substrate friction coefficient between the adhered stick state and the deadhered slip state. 
Here, we consider the dependence of the substrate friction coefficient on both mechanical and chemical signals: 
\begin{equation}
\tau_{\zeta} \frac{d \zeta_i}{d t} = h_{\zeta}(\zeta_i) - A_v h_v(v_i) - (1-A_v ) h_V(V_i), 
\label{eq:dzeta_F_sv_sU}
\end{equation}
where $\tau_{\zeta}$ is the time delay. 
$A_v$ takes a value between 0 and 1, representing the ratio of the mechanical and chemical dependence of the stick-slip transition of the substrate friction. 
See also Fig.~\ref{fig:zeta_function} for the plot of these functions. 

The function $h_{\zeta}(\zeta)$ is defined by
\begin{equation}
h_{\zeta} (\zeta) = - \frac{1}{2} \tanh \Big( \frac{\zeta -\zeta_{\rm stick}}{\epsilon_{\zeta}} \Big) -\frac{1}{2} \tanh \Big( \frac{\zeta -\zeta_{\rm slip}}{\epsilon_{\zeta}} \Big), 
\label{eq:g_zeta}
\end{equation}
where $\zeta_{\rm stick}$ and $\zeta_{\rm slip}$ are the substrate friction coefficients in the adhered stick state and the deadhered slip state, respectively. 
The small parameter $\epsilon_{\zeta}$ indicates the sharpness of the adhesion-deadhesion transition. 
Here, we set the transition sharpness to $\epsilon_{\zeta} = \zeta_{\rm slip} /2$. 
We note that, if there are no change in the substrate friction coefficient, the cell does not exhibit any translational motion. 

If we consider an artificial vesicle or droplet sitting on a substrate, its adhesion strength changes depending on the force acting on it~\cite{Schwarz2013Physics}. 
The term $h_v(v)$ in Eq.~\eqref{eq:dzeta_F_sv_sU} represents this dependence of the cell adhesion to the substrate. 
Here, instead of the force acting on each subcellular element, we presume that the local velocity changes the adhesion strength through
\begin{equation}
h_v(v) = \frac{ ( v/v^* )^2 }{ 1 + ( v/v^* )^2 } -\frac{1}{2},
\label{eq:g_v}
\end{equation}
where $v^*$ is the threshold value. 
The subcellular element tends to adhere to the substrate ($\zeta = \zeta_{\rm stick}$) if the speed is smaller than the threshold value, i.e., $v<v^*$, while the element slips on the substrate ($\zeta = \zeta_{\rm slip}$) if $v>v^*$. 
We set the threshold value to $v^*=1$. 

The formula of the stick-slip transition of the cell-substrate friction depending on the local velocity, i.e., Eq.~\eqref{eq:dzeta_F_sv_sU} with $A_v=1$, was introduced in Ref.~\cite{Barnhart2015Balance}. 
As a result of the balance of the two functions, $h_{\zeta}(\zeta)$ and $h_v(v)$, the substrate friction switches between the stick and slip states with a sharp transition. 

In addition to the mechanical dependence, the adhesion strength of a cell can change depending on its internal chemical conditions~\cite{Schwarz2013Physics}. 
Since the molecular details of cell adhesion are complicated, we assume here that it changes depending on the PIP3 concentration as the actuator elongation: 
\begin{equation}
h_V(V) = \frac{1}{2} \tanh( \sigma_V (V -V^*)).
\label{eq:g_V}
\end{equation}
In Eq.~\eqref{eq:g_V}, $\sigma_V$ stands for the sensitivity, and $V^*$ is the threshold concentration. 
Due to this term, a large value of $V$ prevents strong adhesion if $\sigma_V > 0$, while large $V$ enhances the adhesion if $\sigma_V < 0$. 
However, we are not sure whether PIP3 enhances or diminishes adhesion.

\begin{figure}[t]
\centering
\includegraphics{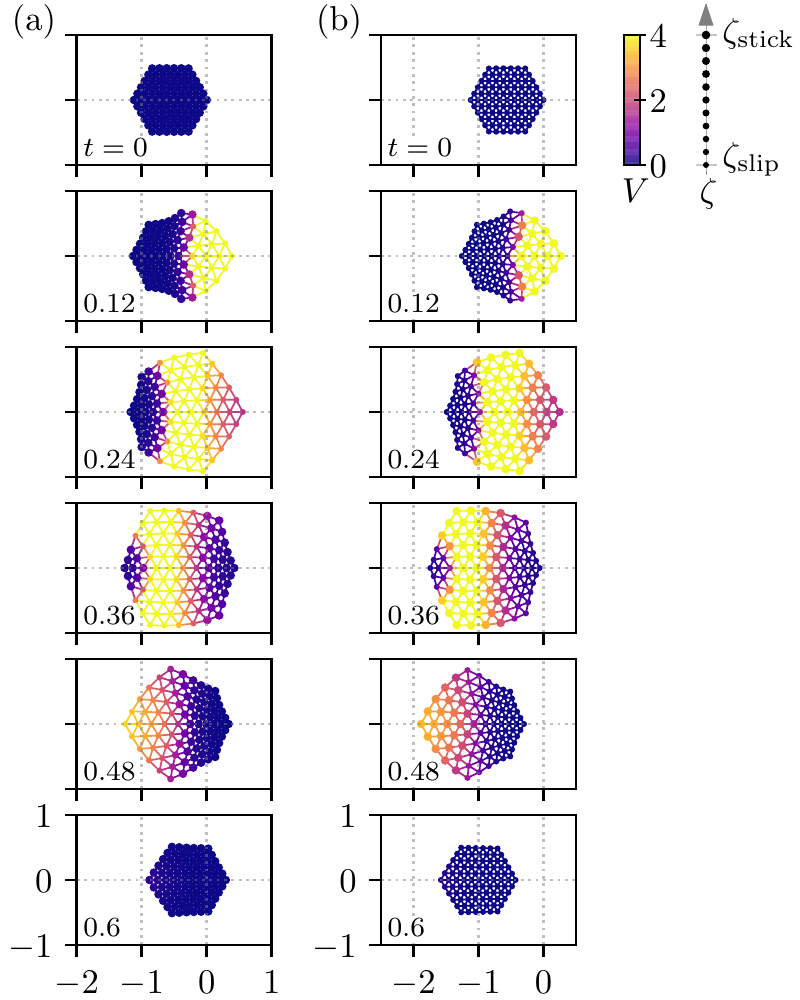}
\caption{
	Cell crawling obtained from Eqs.~\eqref{eq:force_balance}--\eqref{eq:dzeta_F_sv_sU} for different signs of $\sigma_V$: (a) $\sigma_V = 2\pi$ and (b) $\sigma_V = -2 \pi$. 
	The cell for positive $\sigma_V$ crawls in the opposite direction against the traveling chemical wave as shown in panel (a), whereas, for negative $\sigma_V$, it moves in the same direction as the wave as displayed in panel (b).
	The other parameters are set to $A_v=0$ and $\tau_{\zeta}=0.01$. 
	The position of each subcellular element is plotted by a circle whose size and color indicate the value of $\zeta_i$ and $V_i$, respectively. 
	The color of the connecting bonds corresponds to $V_{ij}$. 
	The number in the bottom left corner of each subplot represents the time. 
}%
\label{fig:direct_retrograde_flow}
\end{figure}

\section{Crawling by direct and retrograde waves} \label{sec:direct_retrograde_waves}

To determine the sign of $\sigma_V$, we numerically integrated Eqs.~\eqref{eq:force_balance}--\eqref{eq:g_V} for both positive and negative $\sigma_V$. 
See Appendix~\ref{sec:method} for the details of the simulation. 
Fig.~\ref{fig:direct_retrograde_flow} depicts a time series of snapshots of a crawling cell for $\sigma_V=2\pi$ in Fig.~\ref{fig:direct_retrograde_flow}(a) and $\sigma_V=-2\pi$ in Fig.~\ref{fig:direct_retrograde_flow}(b), respectively.
Here, we set the threshold in Eq.~\eqref{eq:g_V} to $V^*=0.5$ throughout this paper. 

If $\sigma_V$ is positive, the cell moves in the opposite direction to the PIP3 traveling wave, as shown in Fig.~\ref{fig:direct_retrograde_flow}(a). 
Interestingly, however, if the sign of $\sigma_V$ is negative, a qualitatively different result appears: 
namely, the cell starts to move in the same direction as the traveling wave, as displayed in Fig.~\ref{fig:direct_retrograde_flow}(b). 

With respect to the migration direction, the traveling wave in the same direction is called the direct wave, while the one in the opposite direction is referred to as the retrograde wave~\cite{Iwamoto2014The}. 
In this sense, the above crawling motion for positive $\sigma_V$ in Fig.~\ref{fig:direct_retrograde_flow}(a) corresponds to the motion with the retrograde wave, and the one for $\sigma_V<0$ in Fig.~\ref{fig:direct_retrograde_flow}(b) corresponds to the motion with the direct wave. 
Since the experiments in Ref.~\cite{Taniguchi2013Phase} show that cells move in the direction in which PIP3 concentration is increased and thus actin polymerization is enhanced, we set $\sigma_V = 2\pi$ for the rest of this paper.

\begin{figure*}[t]
\centering
\includegraphics{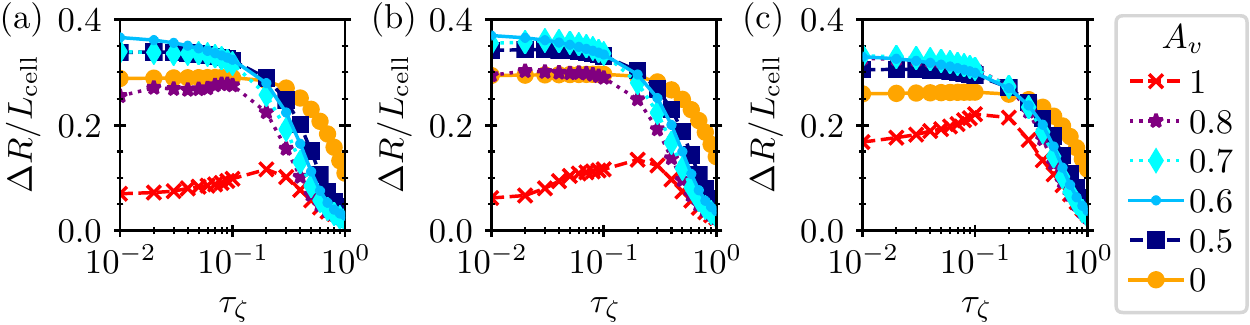}
\caption{
	Normalized migration distance $\Delta R/L_{\rm cell}$ against the time delay $\tau_{\zeta}$ for (a) a hexagonal cell ($L_{\rm cell} = 1$), (b) a circular cell ($L_{\rm cell} = 1$), and (c) a large hexagonal cell ($L_{\rm cell} = 1.4$). 
	The results for several values of $A_v$ are plotted by different lines. 
}%
\label{fig:deltaR}
\end{figure*}

\section{Mechanical vs.\ chemical control of adhesion} \label{sec:adhesion_mechanochemical}

Now, we study the effect of the mechanical and chemical dependence of substrate friction coefficient on cell crawling.
To characterize the cell migration, we measure the migration distance $\Delta R$ of the cell in one cycle, i.e., with one traveling wave, varying the time delay $\tau_{\zeta}$. 
Varying $A_v$ between 0 and 1 tunes the ratio of the mechanical and chemical dependences. 

The results are summarized in Fig.~\ref{fig:deltaR}(a), where different lines correspond to different values of $A_v$ as indicated in the legend. 
Interestingly, the mixing of the mechanical and chemical dependences of the substrate friction may result in larger values of $\Delta R$ than purely mechanical or chemical control. 
In Fig.~\ref{fig:deltaR}(a), $\Delta R /L_{\rm cell}$ reaches approximately 0.36 for $A_v=0.6$ and $\tau_{\zeta}  =0.01$. 
This value, however, is smaller than the crawling of a Dictyostelium cell, which moves its body length in approximately two cycles~\cite{Tanimoto2014A}.

\section{Impact of cellular shape}  \label{sec:cell-shape}

Thus far, we have assumed a hexagonal cell shape, where the structure of the subcellular elements is given by a perfect hexagonal lattice when the cell is at rest. 
However, this structure does not describe real cells, which are instead circular or often of more complicated shapes. 
To elucidate the impact of the cell shape on the crawling motion, we prepare a cell of circular shape as described in Appendix~\ref{sec:Circular}. 
We then perform the simulation and measure the migration distance for different values of $A_v$. 
The results are plotted in Fig.~\ref{fig:deltaR}(b) and are qualitatively the same as those of the hexagonal cell in Fig.~\ref{fig:deltaR}(a).

\section{Impact of cell size}  \label{sec:cell-size}

Now, we study the impact of the size of the cell. 
We prepare a hexagonal cell of size $L_{\rm cell} = 1.4$, which is approximately twice the area of the previous cells. 
We again measure the migration distance $\Delta R$, which is normalized by the cell length $L_{\rm cell}$ to facilitate comparison with the previous results for $L_{\rm cell}=1$. 
The results are summarized in Fig.~\ref{fig:deltaR}(c). 
Qualitatively, the tendency is the same as that of the results in Fig.~\ref{fig:deltaR}(a) for $L_{\rm cell}=1$. 
Namely, the migration distance can be larger if the substrate friction depends both on the mechanical and chemical signals than if it depends on only either one of them.

\section{Random excitation}  \label{sec:random}

\begin{figure*}[t]
\centering
\includegraphics{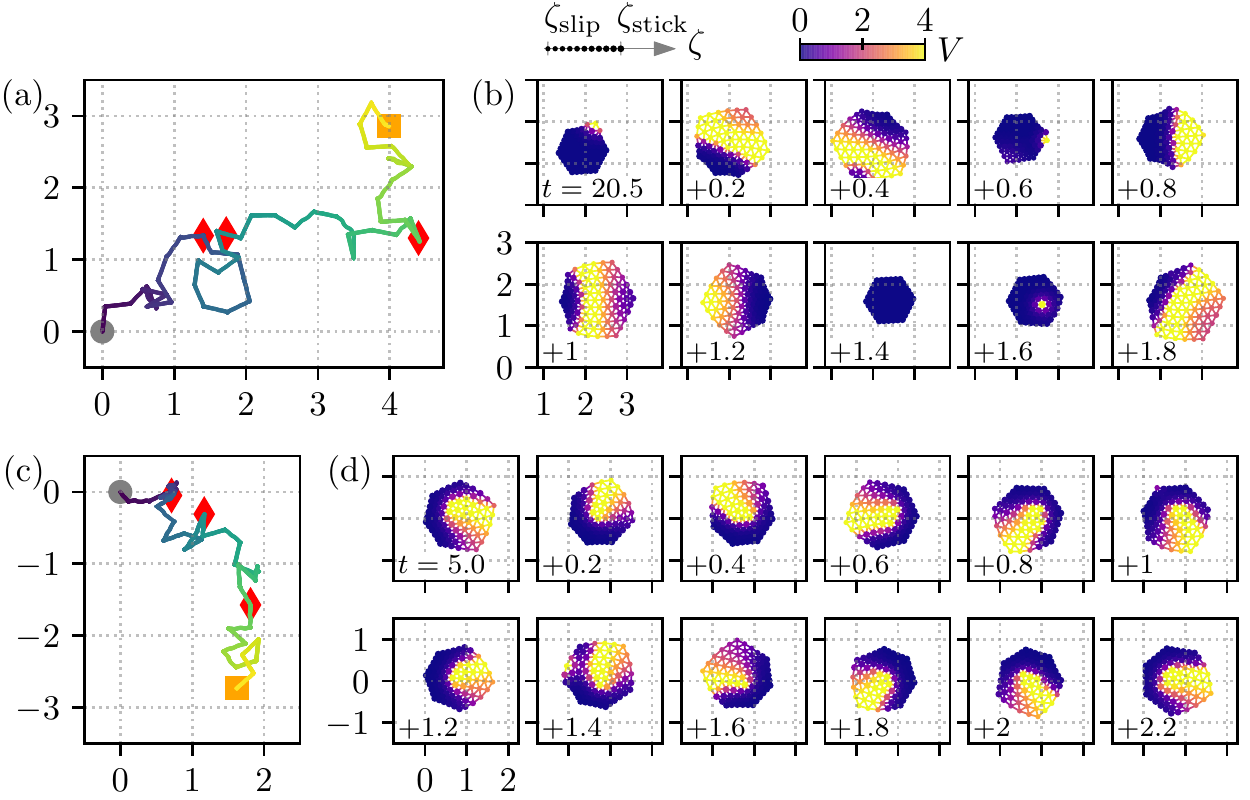}
\caption{
	Cell crawling with randomly-chosen activator element for (a,b) $K_K=5$ and (c,d) $K_K=5.5$. 
	The trajectory of the center-of-mass position is plotted in (a) and (c). 
	The gray circles indicate the initial position at $t=0$, and the red triangles represent the position at every time interval of 10. 
	The orange squares are the final position at time 40. 
	(b,d) Time series of snapshots, (b) where the crawling direction of the cell changes and (d) where the cell undergoes a spinning motion. 
	The numbers at the left bottom corners of the subplots in panels (b) and (d) show the time. 
	The other parameters are the same as in Table~\ref{table:parameters_RD}.
}%
\label{fig:random}
\end{figure*}

In reality, cells change their migration direction over time. 
In our model, we can reproduce such motion by introducing stochasticity, which may originate from, e.g., the complexity of intracellular processes. 
Here, we randomly choose one element in every $t=0.15$ and add to that element the stimulus $(\delta U,\delta V)=( -I_{\rm excite}, +I_{\rm excite})$ with intensity $I_{\rm excite}=0.75$. 

The results are summarized in Fig.~\ref{fig:random}. 
Due to the stochasticity, the cell changes its migration direction frequently, as shown in the trajectory of the center-of-mass position in Fig.~\ref{fig:random}(a). 
From the snapshots of the cell in Fig.~\ref{fig:random}(b), we can see that the migration direction depends on the position at which the chemical wave occurs. 
Because of the excitable nature of the RD equations, a stimulus on the element that has relaxed back to the resting state is more likely to be the origin of the next wave. 
Therefore, a new wave tends to originate from elements that are near the origin of the previous wave. 
As a result, the cell tends to maintain the same migration direction for a time, as shown in Fig.~\ref{fig:random}(a). 
In Figs.~\ref{fig:random}(a) and (b), the parameter values are kept the same as in Fig.~\ref{fig:direct_retrograde_flow}(a). 

If the parameter $K_K$ is slightly increased from 5 to 5.5, the cell changes its migration direction more frequently, as depicted in Fig.~\ref{fig:random}(c). 
Depending on the random stimuli, the cell switches from directional motion to spinning motion as a spiral wave appears, as shown in Fig.~\ref{fig:random}(d). 
The spinning motion is rather stable, but the cell can also switch back to directional motion in response to a stimulus. 
Note that the RD equations maintain their excitable nature at this parameter value.

\section{Traction force multipoles}  \label{sec:traction}

\begin{figure}[t]
\centering
\includegraphics{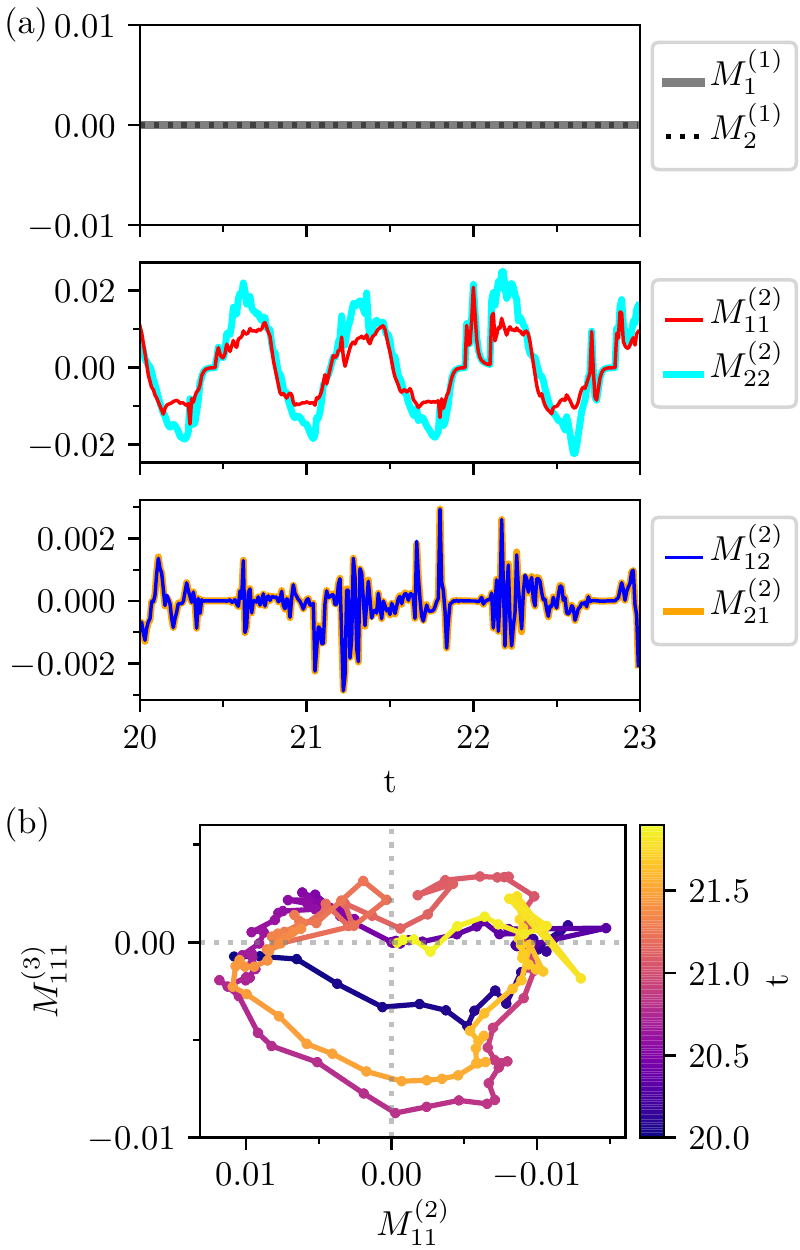}
\caption{
Traction force multipole of the randomly crawling cell in Fig.~\ref{fig:random}(a). 
(a) Times series of each component of traction force monopole ($M^{(1)}_1$ and $M^{(1)}_2$) and the diagonal ($M^{(2)}_{11}$ and $M^{(2)}_{22}$) and off-diagonal components ($M^{(2)}_{12}$ and $M^{(2)}_{21}$) of the traction force dipole. 
(b) Time evolution of traction force dipole $M^{(2)}_{11}$ and quadrupole $M^{(3)}_{111}$. 
The color indicates the time. 
The axis of the $M^{(2)}_{11}$ is inverted, to match the plot in Ref.~\cite{Tanimoto2014A}. 
}%
\label{fig:traction_multipole}
\end{figure}

In many experiments of cell crawling, traction force is measured~\cite{Style2014Traction} since it is of fundamental importance for cell motility. 
Here, we measure the traction force of our model cell. 
Since traction force is the force that a cell exerts on the substrate underneath, it is described by the interaction between the cell and the substrate: $\bm{f}^{\rm traction}_i = \zeta_i(t) \bm{v}_i$. 
We calculate the traction force force multipoles, which are defined as follows, to compare with the experimental result~\cite{Tanimoto2014A}. 

First, the traction force monopole is defined by
\begin{equation}
M^{(1)}_{\alpha} = \sum_i \bm{f}^{\rm traction}_{i,\alpha},
 \label{eq:traction_monopole}
\end{equation}
where $i$ represents the subcellular element $i$ and the summation runs over the entire cell. 
The subscript $\alpha$ indicates spatial component $\alpha = 1,2$. 
Here, the traction force multipoles are calculated in the comoving coordinate on the cell, and thus, $\alpha=1$ and 2 represent the components parallel and perpendicular to the centre-of-mass velocity, respectively, to compare with the experimental results in Ref.~\cite{Tanimoto2014A}. 
In the numerical simulation of the model crawling cell, the traction force monopole is equal to 0, as shown in Fig.~\ref{fig:traction_multipole}, which is consistent with the experimental result~\cite{Tanimoto2014A}. 
This is readily understood from the force balance equation, Eq.~\eqref{eq:force_balance}, and the force-free condition, Eq.~\eqref{eq:force-free}(a). 
That is, the traction force monopole vanishes because of the fact that the inertia is negligibly small for crawling cells on top of the force-free condition. 

The next lowest mode is the traction force dipole, the element of which is defined by
\begin{equation}
M^{(2)}_{\alpha\beta} = \sum_i r_{i,\alpha} f^{\rm traction}_{i,\beta}. 
 \label{eq:traction_dipole}
\end{equation}
On the one hand, from Fig.~\ref{fig:traction_multipole}(a), the diagonal components of the traction force dipole are oscillating around 0. 
Here, note that the positive and negative force dipoles represent the extensile and contractile force. 
In the experiment~\cite{Tanimoto2014A}, only the contractile traction force dipole was observed, which our results fail to reproduce. 
The reason why our model does not reproduce the contractile force dipole is not clear yet. 
One possibility is that the friction coefficient of the protrusion process may be different from that of the contraction process, which are set to the same in our current model. 

On the other hand, the off-diagonal components $M^{(2)}_{12}$ and $M^{(2)}_{21}$ take the same values, indicating that the traction force dipole is symmetric, although such symmetry is not presumed in its definition, Eq.~\eqref{eq:traction_dipole}. 
Such symmetric property of the traction force dipole was also obtained in the experiment~\cite{Tanimoto2014A}.

Now, we consider the meaning of the symmetric property of the traction force dipole that is obtained in our simulation as well as in the experiment. 
Actually, this symmetric property of the traction force dipole indicates the torque-free nature of the cell. 
Here, the torque is defined by
\begin{align}
\bm{T} = \sum_i \bm{r}_i \times \bm{f}^{\rm traction}_i, 
 \label{eq:torque}
\end{align}
which, in a two dimensional space, becomes 
\begin{align}
T = \sum_i ( r_{i,1} f^{\rm traction}_{i,2} -r_{i,2} f^{\rm traction}_{i,1} ) = M^{(2)}_{12} -M^{(2)}_{21} 
 \label{eq:torque_2d}
\end{align}
by using Eq.~\eqref{eq:traction_dipole}. 
Note that, if one describes the force dipole tensor with its invariant, the torque appears as the imaginary part of the eigenvalues.

Finally, in Fig.~\ref{fig:traction_multipole}(b), we plot the time evolution of the traction force dipole $M^{(2)}_{11}$ and the traction force quadrupole $M^{(3)}_{111}$, which was also measured in the experiment~\cite{Tanimoto2014A}. 
Here the traction force quadrupole is defined by
\begin{equation}
M^{(3)}_{\alpha\beta\gamma} = \sum_i r_{i,\alpha} r_{i,\beta} f^{\rm traction}_{i,\gamma}. 
 \label{eq:traction_quadrupole}
\end{equation}
Interestingly, the trajectory in $M^{(2)}_{11}$-$M^{(3)}_{111}$ space shows a counterclockwise rotation, which is qualitatively consistent with the experimental results~\cite{Tanimoto2014A}.

\section{Summary and Discussion} \label{Summary and Discussion}

To summarize, we have constructed a mechanochemical model of a cell crawling on a substrate. 
The mechanical part is described by a subcellular-element model, and the chemical part is described by RD equations. 
To combine them, we introduce two mechanical activities.
One is the actuator of the bond connecting each pair of the subcellular elements, which elongates depending on the intracellular chemical concentration. 
The other is the substrate friction coefficient of each subcellular element, which shows a sharp transition between the adhered stick state and the deadhered slip state. 
We consider the dependence of the substrate friction coefficient on both the local velocity and the intracellular chemical concentration. 
We also introduce a time delay of the substrate friction change. 

By using this model, we clarified that the substrate adhesion dynamics affect how the intracellular force is converted cell crawling motion. 
Depending on the sign of the sensitivity of the substrate friction coefficient to the PIP3 concentration, the model cell exhibited crawling with the retrograde flow or with the direct flow. 
For the former case, which is consistent with experimental observations, our model showed that there is an optimum time delay and that the combined effect of the mechanical and chemical signals on the substrate friction coefficient can increase the migration distance. 
We also investigated the impact of the cell shape and the cell size, which led to qualitatively the same results. 
In addition, we included stochasticity in the RD equations, enabling the cell to change its migration direction and to change its dynamical mode from translational motion to spinning motion. 
Further, we performed multipole analysis of the substrate traction force, which was qualitatively consistent with the experimental results except the contractile nature of the traction force dipole.

Finally, we discuss some possible extensions of our current model. 

\begin{description}
	\item[Contraction process]
	In our current model, the protrusion and contraction processes are both modeled by the actuator elongation, $\ell^{\rm act}_{ij}(t)$, of the bond connecting two subcellular elements. 
	The two processes are distinguished by the sign of $\ell^{\rm act}_{ij}(t)$. 
	In this paper, however, we consider only the protrusion process, i.e., $\ell^{\rm act}_{ij}(t) \geq 0$, which is related to the PIP3 concentration. 
	One reason of this is that the chemical reactions that regulate contraction process are not well understood yet. 
	Therefore, in principle, we can also consider the contraction process by introducing dependence on relevant chemical concentrations. 
	
	\item[Adhesion dynamics]
	The cell adhesion is simply modeled by the switching of the substrate friction coefficient of each subcellular element in our model. 
	However, in real cells, adhesion is mediated by adhesion molecules, which can diffuse and form focal adhesions. 
	To represent these processes of adhesion molecules, we include detailed dynamics of the concentration of adhesion molecules and their diffusion to other subcellular elements. 
	Then, we can discuss detailed structures such as the footstep-like focal adhesion observed for Dictyostelium cells~\cite{Tanimoto2014A}. 
	
	\item[Shape deformation]
	Our model cell shows a lateral expansion with respect to the crawling direction, as shown in Fig.~\ref{fig:direct_retrograde_flow}(a). 
	However, real cells, e.g., Dictyostelium cells, tend to elongate in the direction of motion~\cite{Maeda2008Ordered,Bosgraaf2009The}. 
	One possible reason that our current model fails to reproduce this elongated shape is that actuator elongation depends only on the absolute value of $V_i$. 
	Therefore, this inconsistency may be resolved by, e.g., introducing dependence on the gradient of $V_i$. 
	
	\item[Three dimensions]
	In this paper, we modeled a cell as a two-dimensional network of viscoelastic springs by assuming crawling on a flat substrate. 
	In reality, however, cells are three dimensional object. 
	The extension of our current model to three dimensions is straightforward. 
	
\end{description}

\section{Conclusion} \label{sec:conclusion}

In conclusion, the modeling of crawling cells is still a challenging task due to the complexity of intra- and intercellular processes. 
The force that a cell generates should satisfy the force-free condition, where the total force vanishes. 
To achieve net migration under the force-free condition, the cell needs to break symmetry. 

In our mechanochemical subcellular-element model, the intracellular force acts symmetrically on each pair of subcellular elements; therefore, it naturally satisfies the force-free condition. 
Symmetry breaking occurs due to the switching of the substrate friction coefficient between the adhered stick state and the deadhered slip state. 
Therefore, our model clearly distinguishes intracellular force and external force. 
To control those mechanical activities, we included RD equations representing intracellular chemical reactions. 

The RD equations that we employed in this study were introduced to explain the chemical traveling wave observed within Dictyostelium cells~\cite{Taniguchi2013Phase}. 
However, a number of chemical reactions occur inside a cell, and which chemical reactions are relevant may depend on the phenomena of interest. 
Nevertheless, we believe that our model can provide a basic framework for the future construction of mechanochemical models of crawling cells by replacing the RD equations with suitable ones for each specific phenomenon.

\begin{acknowledgements}%
	This work was supported by the Japan Society for the Promotion of Science (JSPS) KAKENHI (17H01083, 19H14673) grant, as well as the JSPS bilateral joint research projects.
\end{acknowledgements}%

\appendix

\section{Simulation method} \label{sec:method}

We integrate Eqs.~\eqref{eq:force_balance}--\eqref{eq:dzeta_F_sv_sU} numerically by using the Euler method with time increment $\delta t = 10^{-5}$. 
We add the stimulus $(\delta U, \delta V)=( -I_{\rm excite},+I_{\rm excite})$ to Eq.~\eqref{eq:Taniguchi_RD} on one of the subcellular elements, which we call the activator element. 
We set the activator element to the rightmost element denoted by the star in Fig.~\ref{fig:ScEM_schematics}(a), except in the case of random motion, for which an activator element is chosen randomly every time $t=0.15$. 
Since the traveling wave is generated by the stimuli on the activator element, we excite the activator element every $t=1.5$ so that the wave travels to the other edge of the cell within that period, and all the subcellular elements relax back to the resting state when the next wave is generated. 
The intensity of the initial stimulus is set to $I_{\rm excite}=0.75$.

\section{Parameter values} \label{sec:scaling}

The typical values of the mechanical parameters that appear in the model Eq.~\eqref{eq:force_balance} are summarized in Table~\ref{table:scales} and Table~\ref{table:scales_real}. 
The length of the cell at rest is set to $L_{\rm cell} = 1$, and the total number of subcellular elements is set to $N=91$, which gives the total number of bonds as $N_{\rm bond} = 240$, unless otherwise stated. 
For a larger cell, we set $L_{\rm cell} = 1.4$, $N=169$, and $N_{\rm bond} = 462$. 
For a cell consisting of a perfect hexagonal lattice, as in Fig.~\ref{fig:ScEM_schematics}(a), the element distance at rest is set to $\ell_{ij} = L_{\rm cell} /\sqrt{N}$. 
Then, the mean bond length is $\ell_0 = L_{\rm bond} /N_{\rm bond} = L_{\rm cell} /\sqrt{N}$, where $L_{\rm bond}$ is the total bond length. 
The elastic modulus of each bond is set to $\kappa = 10$ and the intracellular dissipation rate to $\xi = L_{\rm cell}^2 /L_{\rm bond}$, which gives the dimensionless parameter $\omega \xi_{\rm cell} /\kappa_{\rm cell} = 0.2 \pi L_{\rm cell}^3$. 
From the mechanical parameter values of typical living cells in Table~\ref{table:scales_real}, the value of this dimensionless parameter is estimated about 0.1--1. 
The substrate friction coefficient at the deadhered slip state is set to $\zeta_{\rm slip} = L_{\rm cell}^2 /N$, so that the dissipation by the substrate friction is on the same order as the intracellular dissipation, $\omega \zeta_{\rm cell} /\kappa_{\rm cell} = \omega \xi_{\rm cell} /\kappa_{\rm cell}$. 
The substrate friction coefficient at the adhered stick state is set to be ten times larger than at the deadhered slip state, i.e., $\zeta_{\rm stick} = 10 \zeta_{\rm slip}$. 
The value of $\sigma$ in $U^{\rm area}$ is set to $\sigma = 10^{-6}$. 

\begingroup
\squeezetable
\begin{table}[tb]
\begin{ruledtabular}
	\caption{%
		Mechanical parameters in the model and their values for the simulation of the hexagonal cell.
	}%
	\begin{tabular}{lc}
		Model parameters & Simulation values \\ \hline
		Diameter of cell at rest, $L_{\rm cell}$ & 1 (1.4 for a larger cell)  \\
		Number of subcellular elements, $N$ & 91 (169 for a larger cell) \\
		Number of bonds, $N_{\rm bond}$ & 240 (462 for a larger cell) \\
		Bond length at rest, $\ell_{ij}$ & $L_{\rm cell} /\sqrt{N}$ \\
		Total bond length at rest, $L_{\rm bond}$ & $\sum_{\langle i,j\rangle} \ell_{ij}=N_{\rm bond} L_{\rm cell} /\sqrt{N}$ \\
		Mean bond length, $\ell_0$ & $L_{\rm bond} /N_{\rm bond}=L_{\rm cell} /\sqrt{N}$ \\
		Substrate friction coefficient & \\
		\hspace{2em} in the slip state, $\zeta_{\rm slip}$ & $L_{\rm cell}^2 /N$ \\
		\hspace{2em} in the stick state, $\zeta_{\rm stick}$ & $10\zeta_{\rm slip}$ \\
		Elastic modulus of each bond, $\kappa$ & 10 \\
		Intracellular dissipation rate, $\xi$ & $L_{\rm cell}^2 /L_{\rm bond}$ \\
		Dimensionless parameter, $\omega \xi_{\rm cell} /k_{\rm cell}$ & $2 \pi \xi_{\rm cell} /k_{\rm cell} T_0=0.2 \pi L_{\rm cell}^3$  \\
	\end{tabular}
	\label{table:scales}
\end{ruledtabular}
\end{table}
\endgroup

\begingroup
\squeezetable
\begin{table*}[tb]
\begin{ruledtabular}
	\caption{%
		Mechanical parameter values for typical living cells and corresponding model parameters.
	}%
	\begin{tabular}{lccc}
		Mechanical parameters & Model & Living cells & References \\ \hline
		Migration speed & &  $\approx 10 \,\mu \rm{m} (\rm{min})^{-1}$ & \cite{Maeda2008Ordered} \\
		Traction stress & & $\approx 100 \,\rm{Pa}$ & \cite{Tanimoto2014A} \\
		Traction force & & $10 \,\rm{nN}$ & (estimated) \\
		Cell-substrate friction coefficient & $\zeta_{\rm cell} = \zeta_{\rm slip} N$ & $1 \,\rm{nN} (\mu \rm{m})^{-1} \rm{min}$ & (estimated) \\
		Elastic modulus & $\kappa_{\rm cell} = \kappa/L_{\rm cell}$ & 10--$100 \,\rm{Pa}$ & \cite{Bausch1999Measurement,Micoulet2005Mechanical} \\
		Elastic constant & $k_{\rm cell} = \kappa/L_{\rm cell}$ & $0.1$--$1 \,\rm{nN} (\mu \rm{m})^{-1}$ & (estimated) \\
		Dissipation rate & $\xi_{\rm cell} = \xi  L_{\rm bond}$ & & \\ 
		Fluid/solid transition time & $\xi_{\rm cell} /k_{\rm cell}$ & $\approx 1 \,\rm{sec}$ & \cite{Kasza2007The} \\
	\end{tabular}
	\label{table:scales_real}
\end{ruledtabular}
\end{table*}
\endgroup

\section{Intracellular RD equations} \label{sec:RD}

Here, we explain the RD equations proposed by Taniguchi et al.~\cite{Taniguchi2013Phase}. 
Taniguchi et al. gathered from their experimental observations of Dictyostelium cells that phosphatidylinositol (3,4,5)-trisphosphate (PIP3) promotes actin polymerization and protrusion of the cellular membrane, and therefore, they considered a signaling pathway around PIP3 including phosphatidylinositol (4,5)-bisphosphate (PIP2), PI3-kinase (PI3K), and phosphatase and tension homolog (PTEN). 
By eliminating the dynamics of PI3K and PTEN, they obtained the following set of RD equations~\cite{Taniguchi2013Phase}:
\begin{align}
\frac{\partial U}{\partial t} = D_U \nabla^2 U +G_U( U, V ),
\cr
\frac{\partial V}{\partial t} = D_V \nabla^2 V +G_V( U, V ),
\label{eq:Taniguchi_dU_dV}
\end{align}
where the reaction terms are defined as
\begin{align}
&
G_U( U, V ) = -\frac{\alpha U V^2}{K_K +\langle V^2 \rangle } +\frac{\beta U V}{K_P +\langle U \rangle } +S -\gamma U,
\cr
&
G_V( U, V ) = +\frac{\alpha U V^2}{K_K +\langle V^2 \rangle } -\frac{\beta U V}{K_P +\langle U \rangle } -\mu V.
\label{eq:Taniguchi_GU_GV}
\end{align}
The global couplings are given by
\begin{align}
\langle U \rangle = \frac{1}{N} \sum_{i} U_i,~
\langle V^2 \rangle = \frac{1}{N} \sum_{i} V_i^2, 
\label{eq:Taniguchi_GlobalCoupling}
\end{align}
where the summation is over all subcellular elements. 
Here, $U_i$ and $V_i$ represent the PIP2 and PIP3 concentrations for subcellular element $i$, respectively. 

Some experiments on Dictyostelium cells~\cite{Taniguchi2013Phase,Arai2010Self-organization,Shibata2012Modeling} revealed that  the actin filaments accumulate around the region where the PIP3 concentration increases, and thus, PIP3 promotes actin polymerization, which induces membrane protrusion. 
Therefore, we assumed that the actuator elongation depends on the PIP3 concentration $V$ as in Eq.~\eqref{eqTaniguchi_ell_ij}.

\begin{figure}[tb]
	\centering
	\includegraphics{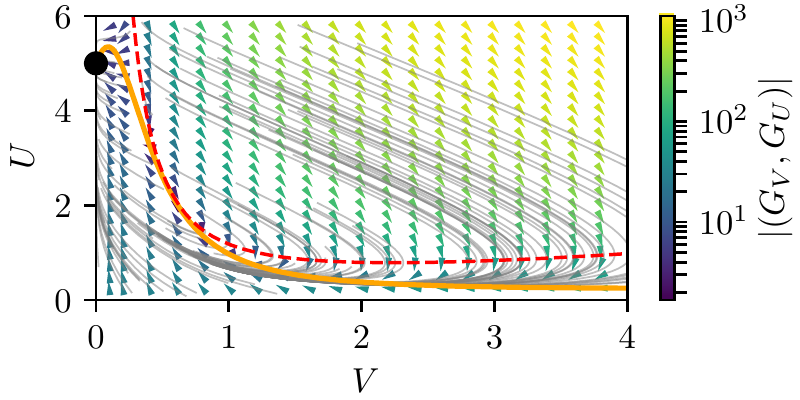}
	\caption{
		The nullcline and the flow field in the $U$-$V$ space of Eqs.~\eqref{eq:Taniguchi_dU_dV}--\eqref{eq:Taniguchi_GU_GV}. 
		The orange solid line and the red dashed line correspond to the nullclines of $G_U(U,V) = 0$ and $G_V(U,V)=0$, respectively. 
		The gray lines show example trajectories starting from various $(U,V)$, whereas the arrow heads and their color show the direction and magnitude of the flow $(G_V(U,V), G_U(U,V))$ at each phase point. 
		The black dot represents the stable fixed point located at $(V,U)=(0,S/\gamma)$. 
	}%
	\label{fig:nullcline}
\end{figure}

The parameters we used in this paper are summarized in Table~\ref{table:parameters_RD}. 
Here, the diffusion coefficient $D_U=0.48$ corresponds to $0.8 \mu \rm{m}^2 \, (\rm{sec})^{-1}$, which is within the range of the experimentally-observed diffusion coefficient of proteins inside a cell near the membrane~\cite{Golebiewska2008Diffusion}. 
We note that the choice of the other parameters is arbitrary, but what is of more importance than their absolute values is the nature of the RD equations, namely, the excitability, which is apparent from the nullclines and the flow field in the $U$--$V$ space, as plotted in Fig.~\ref{fig:nullcline}. 

To integrate Eq.~\eqref{eq:Taniguchi_dU_dV}, we calculate the Laplacian by using the moving particle semi-implicit (MPS) method~\cite{Koshizuka1998Numerical}. 
That is, for the chemical component $c_i = \{ U_i, V_i \}$, the diffusion term is modeled by
\begin{equation}
D_c \nabla^2 c_i = \frac{4 D_c}{\lambda} \sum_{j \neq i} \frac{1}{n_{ij}} (c_j -c_i ) w(r_{ij}),
\label{eq:MPS_laplacian}
\end{equation}
where $n_{ij} = ( n_i + n_j )/2$, and $n_i = \sum_{j \neq i} w( r_{ij} )$ is the number of neighboring elements around element $i$. 
The weight function that we employed is defined by
\begin{equation}
w(r) = \left\{
\begin{array}{ll}
\frac{r_e}{r} -1 & \textrm{for~} 0\leq r < r_e \\
0 & \textrm{for~} r_e \le r
\end{array}
\right.
\label{eq:MPS_w}
\end{equation}
where $r_e$ is the cutoff length, which is set to $4 \ell_0$. 
The normalization factor $\lambda$ is given by $r_e^2 /6$. 
By using this method, we checked whether the traveling and spiral waves are formed in the absence of the mechanical changes. 
To generate the wave, we add the stimulus $(\delta U, \delta V)=( -I_{\rm excite},+I_{\rm excite})$ to Eq.~\eqref{eq:Taniguchi_dU_dV} on the activator element, assuming that the phosphorylation of PIP2 to PIP3 is enhanced for this element.

\begingroup
\squeezetable
\begin{table}[tb]
\begin{ruledtabular}
	\caption{%
		The parameter values of the RD equations in the simulation.
	}%
	\centering
	\begin{tabular}{cc}
		Model Parameters & Values used in simulations \\ \hline
		$D_U$ & 0.48 \\
		$D_V$ & 0.48 \\
		$\alpha$ & 240 \\
		$\beta$ & 90 \\
		$K_K$ & 5 \\
		$K_P$ & 5 \\
		$S$ & 30 \\
		$\gamma$ & 6 \\
		$\mu$ & 30 \\
	\end{tabular}
	\label{table:parameters_RD}
\end{ruledtabular}
\end{table}
\endgroup

\section{Sinusoidal traveling wave} \label{sec:traveling_wave}

\begin{figure*}[tb]
	\begin{center}
		\includegraphics{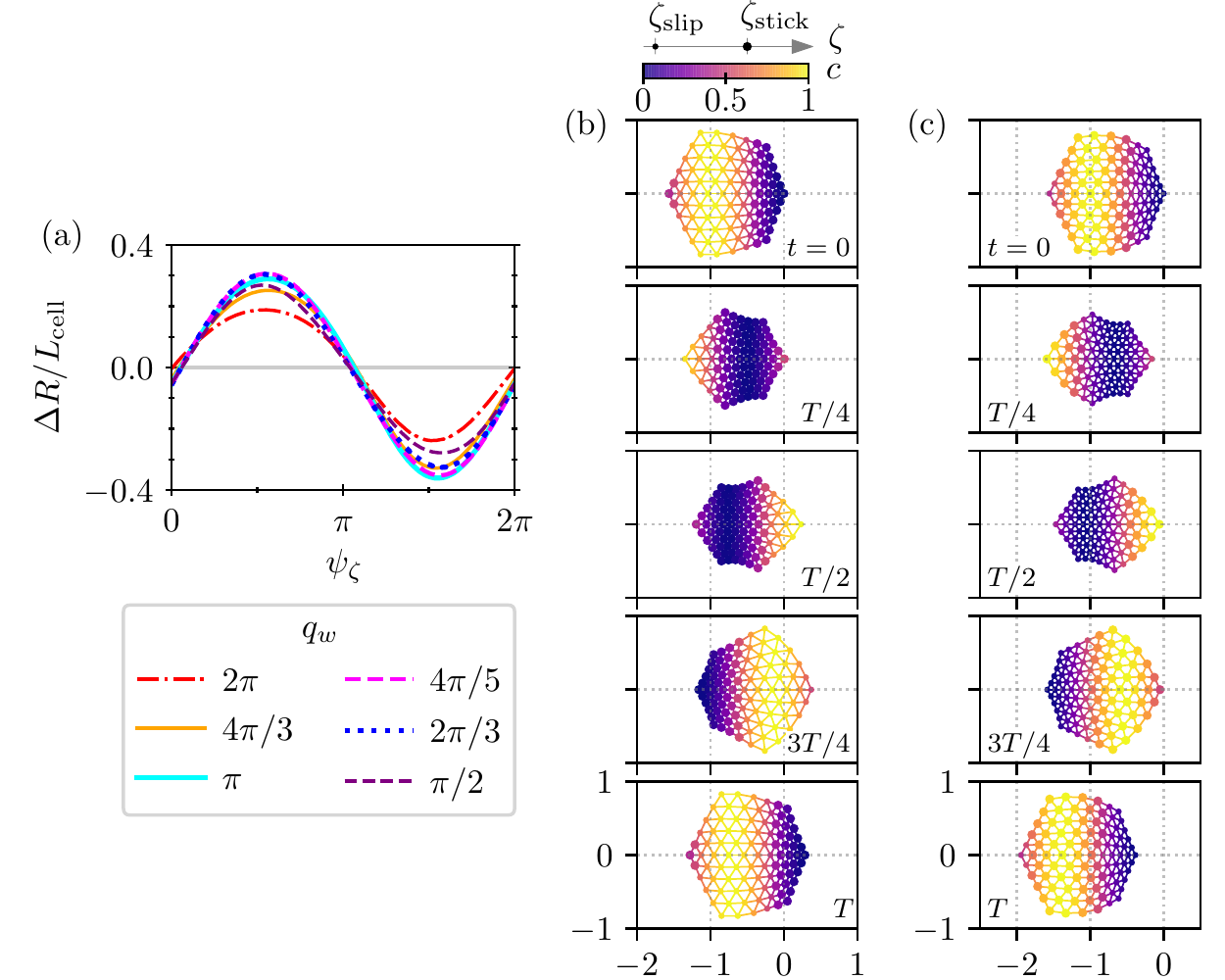}
		\caption{
			Cell crawling with the traveling harmonic wave of intracellular chemical concentration. 
			(a) Migration distance for different values of the wavelength $L_w$ and the phase shift $\psi_{\zeta}$. 
			The displacement in one cycle is measured after several cycles of relaxation. 
			The positive and negative signs of $\Delta R$ correspond to forward and backward motion, respectively. 
			The frequency of the traveling wave is set to $\omega=2 \pi$. 
			[(b) and (c)] Time series of crawling cells for $q_w =\pi$.
			(b) Forward motion for $\psi_{\zeta} = 5\pi/9$ and (c) backward motion for $\psi_{\zeta} = 14\pi/9$. 
			In panels (b) and (c), the numbers show the time, and the color indicates the distribution of the intracellular chemical component $c(t)$. The size of the subcellular elements represents the substrate friction; large and small elements correspond to the adhered stick state and the deadhered slip state, respectively. 
		}%
		\label{fig:DeltaR_sinusoidal}
	\end{center}
\end{figure*}

In this appendix, we show the cell crawling motion induced by a specified intracellular chemical traveling wave. 

For simplicity, we consider only a single intracellular chemical component $c(t)$, which we assume travels periodically inside the cell. 
The chemical component for element $i$ changes as follows:
\begin{equation}
c_i(t) = c_0 ( 1 - \cos \phi_i(t) ),
\label{eq:sin_c}
\end{equation}
where $c_0$ is the maximum concentration, which is set to $c_0=0.5$ so that $0 \le c_i(t) \le 1$. 

For simplicity, we assume a harmonic plane wave that travels in the $x$ direction.
Then, the phase of the chemical concentration $\phi_i(t)$ changes as follows:
\begin{equation}
\phi_i(t) = \omega t -q_w (x_i -x^*),
\label{eq:sin_phi}
\end{equation}
where $\omega$ and $q_w$ are the frequency and the wavenumber of the traveling wave, and
$x^*$ is the $x$ position of the activator element, from which the traveling wave occurs. 

We assume that the cellular activity depends on the value of the intracellular chemical concentration $c_i(t)$. 
The actuator elongation depends on the mean chemical density: 
\begin{equation}
\ell^{\rm act}_{ij}(t) = \ell_c \frac{ c_i(t) +c_j(t) }{2},
\label{eq:sin_ell_ij}
\end{equation}
where $\ell_c$ represents the magnitude of the elongation. 
In addition, the substrate friction coefficient $\zeta_i(t)$ of element $i$ changes depending on $c_i(t)$. 
Here, we assume the substrate friction to be described by a two-state function that switches sharply between the adhered stick state and the deadhered slip state as follows:
\begin{equation}
\zeta_i(t) = \left\{
\begin{array}{ll}
\zeta_{\rm slip} & \textrm{if~} 2m_i \pi < \phi_i(t) -\psi_{\zeta} \le (2m_i +1) \pi \\
\zeta_{\rm stick} & \textrm{if~} (2m_i +1) \pi < \phi_i(t) -\psi_{\zeta} \le 2(m_i+1) \pi
\end{array}
\right.
\label{eq:zeta_i_step}
\end{equation}
where $\psi_{\zeta}$ is a phase shift, $m_i$ is an integer that satisfies $2m_i\pi < \phi_i(t) -\psi_{\zeta} \le 2(m_i+1)\pi$, and 
$\zeta_{\rm stick}$ and $\zeta_{\rm slip}$ are the values of the friction coefficient in the stick and slip states, respectively. 
Here, since the internal motion, i.e., the elongation and the stick-slip transition, is perfectly cyclic, we consider the phase shift instead of a time delay. The two are actually equivalent.

We carried out numerical simulations of Eq.~\eqref{eq:force_balance} together with Eqs.~\eqref{eq:sin_c}--\eqref{eq:zeta_i_step} with the time increment $\delta t =10^{-4}$. 
In the simulation, we choose the element indicated by the star in Fig.~\ref{fig:ScEM_schematics}(a) as the activator element, and we set $q_w>0$ so that the wave travels from right to left. 
The period of the traveling wave is set to the unit time scale, $T=1$, and thus, $\omega = 2\pi$. 
We let the actuator elongate twice as long as its equivalent length: $\ell_c = \ell_{ij}$. 
For these parameters, we vary the wavenumber $q_w$ and the phase shift $\psi_{\zeta}$ and measure the migration distance for one cycle. 

The results are summarized in Fig.~\ref{fig:DeltaR_sinusoidal}.
In Fig.~\ref{fig:DeltaR_sinusoidal}(a), the migration distance for one cycle $\Delta R$ is plotted against the phase shift $\psi_{\zeta}$ for different values of $q_w$, and time series of a typical crawling cell are depicted in Fig.~\ref{fig:DeltaR_sinusoidal}(b) and Fig.~\ref{fig:DeltaR_sinusoidal}(c). 

Fig.~\ref{fig:DeltaR_sinusoidal}(a) shows clearly that the migration direction changes depending on the phase shift $\psi$. 
The positive $\Delta R$ represents forward motion, i.e., migration towards the activator element. 
That is, the cell migrates against the intracellular traveling wave, as depicted in Fig.~\ref{fig:DeltaR_sinusoidal}(b). 
In contrast, the negative $\Delta R$ corresponds to backward motion, where the cell migrates in the same direction as the intracellular traveling wave, as shown in in Fig.~\ref{fig:DeltaR_sinusoidal}(c). 
Therefore, there are maximum and minimum values of $\Delta R$, where the cell achieves maximum migration distance in the forward and backward directions, respectively. 

The migration distance $\Delta R$ also depends on the wavenumber $q_w$.
There is an optimal $q_w$ around $q_w=4\pi/5$ for which the cell can achieve the largest migration distance. 
That is, in this case, the cell is in a fully stretched state.

\section{Circular cell}  \label{sec:Circular}

\begin{figure*}[bth]
	\begin{center}
		\includegraphics{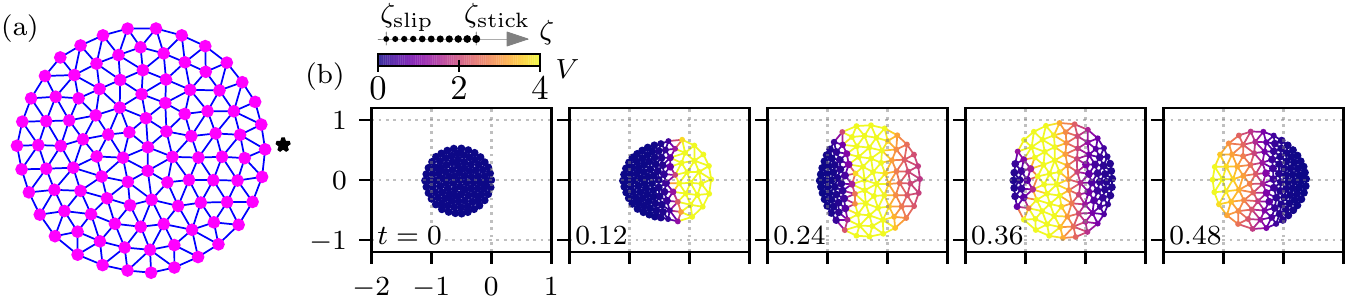}
		\caption{
			Cell with a circular shape. 
			(a) Circular cell at rest. 
			Subcellular elements plotted as magenta circles are connected by viscoelastic bonds indicated by blue lines. 
			The details of the bond are the same as in Fig.~\ref{fig:ScEM_schematics}(b). 
			(b) Time series of crawling circular cells for $A_v=0.6$, $v^*=1$, $V^*=0.5$, and $\tau_{\zeta}=0.01$.
			In each subplot in panel (b), the number in the bottom left corner shows the time, and the color indicates the value of $V(t)$. The size of each subcellular element represents the value of the substrate friction coefficient. 
		}%
		\label{fig:CircularCell}
	\end{center}
\end{figure*}

For cells with a circular shape at rest, we prepared the subcellular elements and the connecting bonds as depicted in Fig.~\ref{fig:CircularCell}(a). 
The length of the cell was set to $L_{\rm cell} = 1$, and the total number of subcellular elements was set to $N=100$. 
The total number of bonds connecting the subcellular elements was then $N_{\rm bond} = 267$. 
We set $\ell_{ij}$ of each bond of the circular cell to the initial element distance, as shown in Fig.~\ref{fig:CircularCell}(a). 
Then, the mean bond length was calculated to be $\ell_0 = 0.105724$. 
The other parameters were kept the same as for hexagonal cells. 

Fig.~\ref{fig:CircularCell}(b) shows an example of a time series of circular cell crawling obtained numerically from Eqs.~\eqref{eq:force_balance}--\eqref{eq:dzeta_F_sv_sU}. 
Here, the activator element is set as the one denoted by the star in Fig.~\ref{fig:CircularCell}(a), which is then stimulated by $(\delta U, \delta V)=( -I_{\rm excite},+I_{\rm excite})$ with $I_{\rm excite}=0.75$ every $t = 1.5$.

\end{document}